\documentstyle[aps,prl,twocolumn]{revtex}
\input epsf
\begin{document}
\draft
\title{Nonequilibrium Fluctuations in Sedimenting Suspensions: A
Dynamical Renormalization Group Theory}
\author{Alex Levine$^1$, Sriram Ramaswamy$^2$, and Robijn Bruinsma$^1$}
\address{$^1$Department of Physics, University of California, Los
Angeles CA 90095, USA\\ $^2$Department of Physics, Indian
Institute of Science, Bangalore 560 012, India}

\date{\today}
\maketitle
\begin{abstract}
A nonlinear two-fluid stochastic hydrodynamical description of
velocity and concentration fluctuations in sedimenting
suspensions is constructed, and analyzed using self-consistent
(SC) and renormalization group (RG) methods. The advection of
particles by velocity fluctuations is shown to be {\em relevant}
in all dimensions $d < 6\/$ . Both RG and SC analyses predict a
strong reduction in the dependence of velocity fluctuations on
system-size $L \/$ relative to the $L^{1/2}\/$ obtained in the
linearized theory of Caflisch and Luke [Phys. Fluids {\bf 28},
785 (1985)]. This is an important step towards resolving a
ten-year old puzzle in the field.
\end{abstract}


Sedimentation \cite{Russel} is at once a rich and complex
phenomenon in colloid science \cite{Hunter,Sood} and a frontier
problem in nonequilibrium statistical mechanics.  The average
sedimentation speed $v_{\mbox{\tiny{sed}}}\/$ is determined by
balancing the driving force (gravity) against the dissipative
force (viscous drag). Despite the long-ranged ($1/r\/$) nature
of the hydrodynamic interaction between the colloids, an adequate
mean-field theory \cite{Batchelor} of $v_{\mbox{\tiny{sed}}}\/$
is available for low concentrations. There is, however, no
satisfactory theory of the {\em fluctuations} of the velocity
and concentration fields in a steadily settling many-particle
suspension, even at zero Reynolds number.  Experiments
\cite{Tory,Nicolai1,Xue,Rutgers} show clear nonthermal
fluctuations in the particle motion, with the character of
random stirring on a scale substantially larger than the
particle size. The fluctuations $\delta v \/$ in the settling
speed are of the same order as the mean $v_{\mbox{\tiny{sed}}}$.
Caflisch and Luke (CL) \cite{Caflisch} showed in a {\em
linearized} hydrodynamic analysis that, for a sedimenting system
of linear dimension $L\/$, the assumption of purely {\em random}
local concentration fluctuations led to long-ranged velocity
fluctuations with $\delta v \sim L^{1/2}\/$.  Experiments,
however, find {\em no} dependence of $
\delta v $ on $L$ \cite{Nicolai1,Xue,Ladd}. In this letter we {\em
assume} a source of randomness at small scales, the origin of
which we comment on briefly below, and examine its consequences
at large scales. In detail, we use a dynamical renormalization
group (DRG), general scaling arguments, as well as perturbative
self-consistent methods to analyze the advection of
concentration fluctuations in steady-state sedimentation by the
velocity fluctuations which they themselves produce. The DRG
method (see Ma and Mazenko \cite{Ma}), was applied to the
closely related problem of the randomly stirred fluid by
Forster, Nelson, and Stephen (FNS) \cite{Forster}. Scaling and
self-consistent approaches give a family of possible results,
the simplest and most physically appealing of which accounts for
the experimentally observed lack of size-dependence. The DRG
results, which are to first order in $\epsilon = 6 - d\/$ where
$d\/$ is the dimension of space, present a clear trend towards
suppressing the size dependence, but simply setting $\epsilon =
3\/$ in these first-order calculations leaves a residual size
dependence, albeit much weaker than the $L^{1/2}\/$ of CL
\cite{Caflisch}. 

We coarse-grain the dynamics on scales comparable to the
random-stirring scale $\ell_s\/$. This obliges us to use a {\em
stochastic} description with a noise source and an effective
bare diffusivity or hydrodynamic dispersion coefficient
\cite{Brady} $D_0 \sim v_{\mbox{\tiny{sed}}} \ell_s\/$. 
We then calculate quantities such as the velocity fluctuation on
scales $\gg \ell_s\/$ using the following two-fluid hydrodynamic
description of sedimentation along ${\hat{\bf z}}\/$. Our 
coupled equations for fluctuations {\bf v} and $c\/$ in the
velocity and concentration fields respectively, at zero Reynolds
number, are 
\begin{equation}
\label{main}
\frac{\partial c}{\partial t} + \lambda {\bf v} \cdot {\bf
\nabla} c = D_o \nabla^2 c + \theta({\bf r},t) 
\end{equation}
\begin{equation}
\label{main2}
\eta \nabla^2 v_i  =   m_R g \delta c P_{iz}\\
\end{equation}
Here the pressure field has been eliminated by imposing
incompressibility via the transverse projector $P_{ij} =
\delta_{ij} - \nabla_i \nabla_j (\nabla^2)^{-1}\/$.
Equation \ref{main} is the advection-diffusion equation for a
scalar field with $ \lambda = 1$. The bare collective
diffusivity  $D_o\/$ and the Gaussian noise term
$\theta\/$, with zero mean and 
\begin{equation}
\label{corr}
\langle \theta({\bf r},t) \theta({\bf 0},0) \rangle = N_o
\nabla^2 \delta({\bf r}) \delta(t) 
\end{equation}
have the same origin --- in general a combination of thermal
fluctuations and hydrodynamic dispersion.  If the noise is of
purely chaotic origin, as in Ladd's athermal simulation
\cite{Ladd}, then $N_o$ and $D_o$ are parameters
which must be determined either experimentally or by the
analysis of numerical simulations.  For purely thermal noise,
the ratio $N_o/D_o$ is fixed by the fluctuation dissipation
theorem (FD) \cite{FDT} while $ D_o \sim k_B T/\eta a $ with
$\eta $ the solvent viscosity and $a$ the radius of the
colloids. Eq.~\ref{main2}, which expresses the balance between
the driving by gravity and dissipation by viscosity,  
describes how concentration fluctuations produce
velocity fluctuations with $m_R g\/$ the buoyancy-reduced
colloidal weight. In Eqs.~\ref{main} -\ref{main2} other possible
nonlinear terms, e.g. those arising from concentration-dependent
mobilities, are readily shown to be subdominant relative to the 
advective nonlinearity $ \nabla. {\bf v} c\/$ \cite{future}.  

If we neglect the nonlinear term ( i.e. set $\lambda = 0$ ),
then we can solve the equations to find the result of
\cite{Caflisch}: 
\begin{equation}
\label{linear}
\langle v_z^2 \rangle_L = 3\langle {\bf v}_\perp^2 \rangle_L
\propto \frac{N_o}{D_o} L 
\end{equation}
The nonlinear term causes these induced velocity fluctuations to
advect further the concentration field. Since flow advection is
well known to distort and suppress fluctuations \cite{Onuki},
the ultimate size-dependence should be weaker than that predicted
by Eq.~\ref{linear}.

\begin{figure}
\epsfxsize=8cm
\centerline{\epsfbox{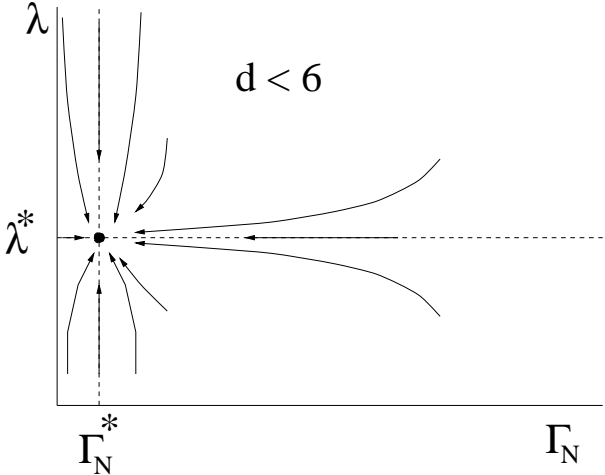}}
\label{nofdt}
\caption[]{The renormalization group flow below the upper
critical dimension in the coupling constant ($\lambda$), noise
anisotropy ($\Gamma_N$) plane.  The IR stable fixed point which
violates the fluctuation - dissipation theorem ($\Gamma_N^\star,
\lambda^\star$) is represented by the black dot at the
intersection of the dashed lines representing the fixed point
values of $\lambda$ and $\Gamma_N$.  } 
\end{figure}
Let us summarize first the simple scaling and self-consistent
arguments.  Note first that within the above linear treatment
the dynamic exponent relating spatial and temporal rescaling is
$z = 2\/$ and the structure factor for concentration
fluctuations $S(k) = <|c_k|^2> \neq 0\/$ for wavenumber $k
\rightarrow 0\/$. The nonlinearity $\nabla . c {\bf v}\/$, whose
advective form guarantees that there will be no graphical
corrections to it under renormalization, alters these
conclusions dramatically. Rescaling (isotropically for
simplicity) $r \rightarrow b r, \, t \rightarrow b^z t, \, c
\rightarrow b^{\chi_c} c, \, v \rightarrow b^{\chi_v}
v\/$, normalizing to retain the form of the first term in
(\ref{main}), and insisting that the coefficient of the
nonlinearity remain unchanged, yields $\chi_c + z +1 = \chi_v +
z - 1 = 0\/$.  We can identify the upper critical dimension
$d_c\/$ above which the nonlinearity is irrelevant by noting
that for $d \geq d_c\/$, linearized behaviour should obtain:
$\chi_c = -d/2, \, z=2\/$. This gives $d_c = 6\/$, which is some
indication of why a reliable answer in three dimensions is going
to be hard to find.  The same results can obtained from a
one-loop self-consistent calculation of the renormalized
diffusivity $D(k,\omega)\/$ and noise spectrum $N(k,\omega)\/$
\cite{future}. In either case, if we {\em assume} that there is
no singular renormalization (except possibly logarithms) of the
noise spectrum \cite{Frey}, we get, for $d < 6\/$, $z = d/3\/$,
and $S(k) \sim 1/D(k) \sim k^{2-d/3}\/$.  In three dimensions
this is equivalent to the result $S(k) \sim k\/$ of \cite{Koch}.
While this last is but one of a family of possible results that
arise out of the scaling analysis, it has a particular physical
appeal, which we discuss at the end of this Letter. 

\begin{figure}
\epsfxsize=8cm
\centerline{\epsfbox{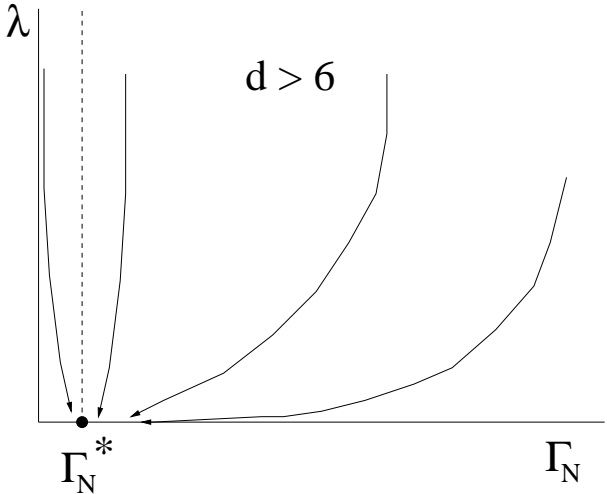}}
\label{fdt}
\caption[]{The renormalization group flow above the upper critical 
dimension in the coupling constant ($\lambda$), noise anisotropy 
($\Gamma_N$) plane.  The IR stable fixed point is that describing the 
free theory with a finite shift in the anisotropy parameter 
$\Gamma_N = N_\perp/N_z$.  This fixed point model obeys the 
fluctuation - dissipation theorem. } 
\end{figure}

We turn now to the DRG calculation \cite{Ma,Forster} of the
renormalization of the dynamical parameters $N, D\/$, and
$\lambda\/$.  The RG recursion relations to one loop order near
the upper critical dimension $ d = 6\/$ are 
\begin{eqnarray}
\label{RG1}
&&\frac{d \bar{\lambda}}{d \ell}  =  \bar{\lambda}(\ell) \times  
\nonumber
\\
&&\left[
\frac{1}{2} \epsilon - 10^{-4} \bar{\lambda}^2(\ell) \left\{ 1.0
+ 9.4 \Gamma_N(\ell) - 3 \Gamma_N^2(\ell) \right\} \right] \\ 
\label{RG2}
&&\frac{d \Gamma_N}{d \ell}  =  
2.62 \times 10^{-6} \bar{\lambda}^2(\ell) \times  
\nonumber
\\
&&\left[ 1 - 5.6 \Gamma_N(\ell) - 163.8
\Gamma_N^2(\ell) - 231 \Gamma_N^3(\ell) \right] \\ 
\label{RG3}
&&\frac{d N}{d \ell}  =  - 10^{-4} N(\ell) \bar{\lambda}^2(\ell) \times 
\nonumber
\\
&&\left[ 0.55 + 5.5 \Gamma_N(\ell) - 5.5 \Gamma_N^2(\ell) \right]
\\ 
\label{RG4}
&&\frac{d \Gamma_D}{d \ell} =  - 6.72 \times 10^{-5}
\bar{\lambda}^2(\ell) \Gamma_D(\ell) \times  
\nonumber
\\
&&\left[ 1 + 9.63
\Gamma_N(\ell) \right] 
\end{eqnarray}
We have allowed explicitly for different noise strengths
$N_z\/$ and $N_{\perp}\/$ for wavevectors along and normal to
$\hat{\bf z}\/$ respectively. Here, the dimensionless coupling
constant is $ \bar{\lambda} = \frac{\lambda
N_z^{1/2}}{D_z^{1/2}}$ , and $ \epsilon = d - 6$, while
$\Gamma_N \equiv N_\perp/N_z$ and $\Gamma_D \equiv D_\perp/D_z$
are the anisotropy factors of respectively the noise and
diffusion constants.  We selected the dynamical exponent such
that $D_z$ has trivial scaling.  Figs.\ref{nofdt} and \ref{fdt} show 
the resulting RG flow lines for $ d > 6$ and $d<6$ respectively. For $d>6$,
the nonlinear coupling constant $\bar{\lambda}(\ell) $ scales to
zero for large $\ell$.  Linearized hydrodynamics is applicable
and $ \langle v^2 \rangle_L \propto L^{4 - d} + \mbox{const} $
in agreement with Eq.~\ref{linear}.  For $ d< 6 $, on the
other hand, we find a stable fixed point
($\bar{\lambda}^{\star}, \Gamma_N^{\star} $).  The noise
is highly anisotropic at the fixed point ($\Gamma_N^{\star}
\approx 0.0608 $), while $\bar{\lambda}^{\star}
\propto \sqrt{\epsilon} $.  The noise amplitude $ N(\ell)$
scales to zero according to Eq.~\ref{RG3}, as does the diffusion
constant anisotropy ( Eq.~\ref{RG4}). This means that the ratio
$N(\ell)/D(\ell) $ of the noise to the diffusion constant scales
to zero at the fixed point signalling the absence of an FD
theorem at large lengthscales below $ d = 6$, irrespective of
the origin of the noise.  There is a FDT respecting fixed point
which is unstable to any bare deviation of the noise from its
FDT determined value.  The linear theory, which in effect
assumes an FD theorem, thus overestimates the magnitude of
concentration fluctuations at large lengthscales.  This
flow-induced noise-suppression is highly anisotropic.  This is
already clear from the fact that $\Gamma^{\star}_N $ is a small
number, so concentration fluctuations with wavevectors along the
sedimentation direction are much larger than concentration
fluctuations perpendicular to the sedimentation direction.  A
similar conclusion can be drawn from the dynamical scaling
exponents.  The relaxation rate $ \omega(q_z) $ for
concentration fluctuations {\em along} the $ \hat{z}$ direction
depends on the wavevector $q_z$ as $ \omega(q_z) \propto
q_z^{z_\|} $, with $z_\|$ the corresponding scaling exponent,
while in the perpendicular direction $ \omega({\bf q}_\perp)
\propto |{\bf q}_\perp|^{z_\perp} $.  If we use $\epsilon = 3 $
in Eq.~\ref{RG1} we find $z_\| \cong 0.91 $ while $z_\perp \cong
1.93$. In-plane relaxation of concentration fluctuations is
thus nearly purely diffusive while out-of-plane relaxation is
much faster than diffusive. 
To make quantitative predictions we use the dynamical scaling relations
\begin{eqnarray}
\label{Dam1}
N_{z,\perp}(q_{z,\perp}, \omega) & = & q_{z,\perp}^\alpha
F_{z,\perp}(\omega/q_{z,\perp}^{z_{\|,\perp}}) \\ 
\label{Dam2}
D_{z,\perp}(q_{z,\perp}, \omega) & = &
q_{z,\perp}^{\beta_{z,\perp}
}G_{z,\perp}(\omega/q_{z,\perp}^{z_{\|,\perp}})
\end{eqnarray}
where $ \alpha = -2 + z_{\|} + \gamma,\, \beta_z = -2 + z_\|, \,
\beta_\perp = -2 + z_\perp$ with $\gamma = 0.83$ for $ \epsilon
= 3$.  The scaling functions $ F_{z,\perp}(x)$ and
$G_{z,\perp}(x)$ are all finite at $ x= 0$ but for large $x$,
they behave as power laws: 
\begin{eqnarray}
\lim_{x \rightarrow} F_{z,\perp}(x) & \propto & x^{\alpha/z_{\|,\perp}} \\
\lim_{x \rightarrow} G_{z,\perp}(x) & \propto &
x^{\beta_{z,\perp}/z_{\|,\perp}} 
\end{eqnarray}
The full velocity fluctuation spectrum is then :
\begin{eqnarray}
&&\langle |v_z^2({\bf q},\omega) | \rangle  = 
\nonumber
\\
&\Delta&
\frac{q_\perp^4}{q^8} \left[ \frac{ N_z({\bf q},\omega) q_z^2 +
N_\perp({\bf q},\omega) q_\perp^2 }{\omega^2 + \left(
D_z({\bf q},\omega) q_z^2 + D_\perp({\bf q},\omega) q_\perp^2
\right)^2} \right] 
\\ 
&&\langle |{\bf v}_\perp^2({\bf q},\omega) |
\rangle  =  
\nonumber
\\
&\Delta &
\frac{q_\perp^2 q_z^2}{q^8} \left[ \frac{
N_z({\bf q},\omega) q_z^2 + N_\perp({\bf q},\omega) q_\perp^2
}{\omega^2 + \left( D_z({\bf q},\omega) q_z^2 +
D_\perp({\bf q},\omega) q_\perp^2 \right)^2} \right] 
\end{eqnarray}
with $\Delta = \left(m_R g/\eta \right)^2 $.  Integrating over
${\bf q} $ and $\omega$ while using the dynamical scaling
relations given by Eqs.~\ref{Dam1},\ref{Dam2} gives our final
result for the size-dependence of the velocity fluctuations in $ d = 3$.
\begin{eqnarray}
\label{resultA}
\langle |v_z|^2 \rangle_L & \propto & L^{1 + (z_\perp - z_\|)/2 - \gamma} \\
\label{resultB}
\langle |{\bf v}_\perp|^2 \rangle_L & \propto & L^{1  - \gamma}
\end{eqnarray}
The trend towards a substantial reduction of the size-dependence
is gratifying. If we simply extrapolate our first-order
in $\epsilon\/$ results to $d= 3$ we find $ \langle |v_z|^2
\rangle_L \propto L^{0.6} $ and $\langle |{\bf v}_\perp|^2 \rangle_L
\propto L^{0.17} $. Comparing Eqs.~\ref{resultA} and
\ref{resultB} with experiment we first note that velocity
fluctuations diverge considerably more weakly than linearly
along the $z$ direction and hardly at all along the
perpendicular direction so Eqs.~\ref{resultA} and
\ref{resultB} are closer to the observed size dependence than
Eqs.~\ref{linear} and \ref{linear}. While experiments report no
size dependence at all, our theory really is only
reliable near $ d= 6$ and a naive extrapolation of a one loop
DRG calculation to $ d= 3$ is questionable.  Another interesting
method of comparison between experiment and the present theory
is in the anisotropy factor $\langle |{\bf v}_\perp|^2 \rangle/
\langle |v_z|^2 \rangle $.  In the linearized theory this
quantity is equal to $1/3$ (Eq.~\ref{linear}) while the
non-linear theory predicts that this ratio vanishes as
$L^{-0.43}$ with system size (Eqs.~\ref{resultA},
\ref{resultB}).  Ladd's simulation gives $0.12$ for this ratio
and recent experiments $0.27$ and $0.26$
\cite{Nicolai1,Chaikin}.  All of these measurements thus suggest
that the ratio is indeed less than $1/3$.  

Before closing, let us note that the na\"{i}ve self-consistent
results in $d = 3\/$ can also be obtained by dimensional
arguments reminiscent of those used in the theory of homogeneous
isotropic turbulence \cite{Tennekes}, but with the
assumption of a scale-independent {\em velocity} rather than
dissipation.  $z = 1\/$ or equivalently $D(k) \sim 1/k\/$ means
that the diffusivity at a length scale $k^{-1}\/$ is given by
that length times a constant which must on dimensional grounds
have the units of velocity. In the absence of external, thermal
noise, the only candidate for this constant is
$v_{\mbox{\tiny{sed}}}\/$ itself. Making the reasonable guess
that {\em all} physical quantities, in the $Pe \rightarrow
\infty,\, Re \rightarrow 0\/$ limit, must then be built from
$v_{\mbox{\tiny{sed}}}$ and wavenumber leads immediately to
$D(k) \sim v_{\mbox{\tiny{sed}}}/k, \, \delta v
\sim v_{\mbox{\tiny{sed}}}\/$ and the mean-square displacement
of particles $\sim v_{\mbox{\tiny{sed}}}^2 t^2\/$
upto logarithmic factors and volume-fraction dependent
coefficients.

Finally, we summarize the achievements of this work. First,
starting from the correct, coarse-grained, nonlinear stochastic
hydrodynamic equations, which in principle are a complete
description of the problem, we have shown clearly that the
origin of the disagreement between theory \cite{Caflisch} and
experiments \cite{Nicolai1,Xue} lies in the neglect of an
important nonlinearity, {\em viz.}, the advection of
concentration fluctuations by the fluctuations in the flow.  Our
work thus emphasizes that both the renormalized sedimentation
speed \cite{Batchelor} and fluctuations therein should be
calculated in parallel with, and not subject to assumptions
about, the structure factor, and shows that stochastic models
are a way of doing this.  Second, our identification of $d =
6\/$ as the upper critical dimension for the problem shows that
the effects of this nonlinearity are likely to be (a) very large
and (b) rather difficult to calculate analytically in three
dimensions. Third, our dynamic renormalization group
calculations in an expansion in $\epsilon = 6 - d\/$ indicate a
clear trend in the direction of reduced system-size dependence.
Finally, with one additional assumption, we can use
self-consistent and dimensional arguments which give results
which imply at most a logarithmic size-dependence, and make
further predictions for experiment.  A reliable calculation
directly in three dimensions remains the main challenge for
theory.

We would like thank Maarten Rutgers and Paul Chaikin for
communicating unpublished results and for useful discussions.
We would also like to thank J. Brady, D. Durian, E. Frey, E.
Herbolzheimer, R. Pandit, and J. Rudnick for useful discussions.
S.R.  thanks F. Pincus and C. Safinya and the Materials Research
Laboratory, UCSB (NSF DMR93-01199 and 91-23045), as well as the
ITP Biomembranes Workshop (NSF PHY94-07194) for partial support.
A. L.  acknowledges support by an AT\&T Graduate Fellowship.

\end{document}